\documentclass[a4paper,11pt]{article}
\pdfoutput=1 

\usepackage{jheppub} 

\usepackage[T1]{fontenc} 
\usepackage{color}
\usepackage{graphicx}
\usepackage{textcomp}
\usepackage{amsmath,amssymb}
\usepackage{enumerate}
\usepackage{multirow}
\usepackage{gensymb}
\usepackage{hyperref}
\newcommand{\mathsym}[1]{{}}

\newcommand{\ba}{\begin{array}} 
\newcommand{\ea}{\end{array}}
\newcommand{\be}{\begin{equation}}
\newcommand{\ee}{\end{equation}}
\newcommand{\beqa}{\begin{eqnarray}} 
\newcommand{\eeqa}{\end{eqnarray}}
\def\321{$SU(3)\times SU(2)\times U(1)$}

\newcommand{\Upmns}{U_{\rm PMNS}}

\newcommand{\group}[1]{\textlbrackdbl #1\textrbrackdbl}

\def\et{\eta}

\title{Residual $Z_2$ symmetries and leptonic mixing patterns from finite discrete subgroups of $U(3)$}

\author[a]{Anjan S. Joshipura,}
\author[b]{Ketan M. Patel}

\affiliation[a]{Physical Research Laboratory,\\ Navarangpura, Ahmedabad 380 009, India}
\affiliation[b]{Indian Institute of Science Education and Research, Mohali,\\ Knowledge City, Sector  81, S A S Nagar, Manauli 140 306, India}

\emailAdd{anjan@prl.res.in}
\emailAdd{ketan@iisermohali.ac.in}

\abstract{We study embedding of non-commuting  $Z_2$ and $Z_m$,  $m\geq 3$ symmetries in discrete subgroups (DSG) of $U(3)$ and analytically work out the mixing patterns implied by the assumption that $Z_2$ and $Z_m$ describe the residual symmetries of the neutrino and the charged lepton mass matrices respectively. Both $Z_2$ and $Z_m$ are assumed to be  subgroups of a larger discrete symmetry group $G_f$ possessing three dimensional faithful irreducible representation. The residual symmetries predict the magnitude of a column of the leptonic mixing matrix $U_{\rm PMNS}$ which are studied here assuming $G_f$ as the DSG of $SU(3)$ designated as type C and D and large number of DSG of $U(3)$ which are not in $SU(3)$. These include the known group series $\Sigma(3 n^3)$, $T_n(m)$, $\Delta(3 n^2,m)$, $\Delta(6n^2,m)$ and $\Delta'(6 n^2,j,k)$. It is shown that the predictions for a column of $|U_{\rm PMNS}|$ in these group series and the C and D types of groups are all contained in the predictions of the $\Delta(
6 N^2)$ groups for some integer $N$. The $\Delta(6N^2)$ groups therefore represent a sufficient set of $G_f$ to obtain predictions of the residual symmetries $Z_2$ and $Z_m$.}

\begin{document} 
\maketitle
\flushbottom

\section{Introduction}
\label{intro}
Discrete subgroups (DSG) of $SU(3)$ and $U(3)$ have been extensively used as flavour symmetries with a view to understand the observed leptonic mixing patterns, see \cite{Altarelli:2010gt,Altarelli:2012ss,Smirnov:2011jv,King:2013eh,Ishimori:2010au} for reviews. Many of these approaches are based on two observations: (a) the  mass matrix $M_\nu$ for the  non-degenerate Majorana neutrinos and a Hermitian combination $M_lM_l^\dagger$ of the charged lepton mass matrix are always invariant under residual symmetries $G_\nu \in Z_2\times Z_2$ and $G_l\in Z_n\times Z_m\times Z_p$ respectively and (b) the unitary matrices $U_{G_\nu}$, $U_{G_l}$ diagonalizing $G_\nu$, $G_l$ respectively also diagonalize the corresponding mass matrices. Thus the forms of the symmetry generators contained in  $G_\nu$, $G_l$ are sufficient to determine the leptonic mixing matrix \cite{Lam:2007qc,Lam:2008rs,Lam:2008sh,Lam:2012ga,Lam:2011ag}. The former can be determined group theoretically by requiring that $G_\nu$ and $G_l$ are subgroups 
of a larger group $G_f$.

The predictions of the above mentioned approaches have been extensively studied \cite{deAdelhartToorop:2011re,Toorop:2011jn,Hu:2012ei}. They have been numerically worked out \cite{Parattu:2010cy,Hagedorn:2013nra,Holthausen:2013vba,Talbert:2014bda} for a large number of DSG of $U(3)$ and the group series $\Delta(6 n^2)$ \cite{King:2013vna} under the assumption of $G_\nu=Z_2\times Z_2$ and $G_l=Z_m$, $m\geq 3$.
General classification of all the possible mixing patterns under the above assumptions is also presented in \cite{Fonseca:2014koa} from a different point of view. The virtue of the above approach is its predictivity since the matrices diagonalizing the generators of $Z_2\times Z_2$ and $Z_m$, $m\geq 3$ (with unequal eigenvalues) are completely fixed apart from over all diagonal phase matrices. A less conservative approach is to assume that only a $Z_2$ subgroup of the full  neutrino residual symmetry $Z_2\times Z_2$ resides in the flavour group $G_f$. This allows additional phases and a mixing angle which do not get determined from group theoretical considerations. This approach is desirable since it is found that only a handful of $G_f$ containing the  full $Z_2\times Z_2$ are actually able to explain all the three leptonic mixing angles at the leading order \cite{Holthausen:2012dk,Yao:2015dwa}. Moreover, there exist a large number of DSG of $SU(3)$ (those with order not divisible by 4 but divisible by 2) 
which contain $Z_2$ but not the full Klein group $Z_2\times Z_2$. Predicted mixing patterns in this approach are also studied numerically in \cite{Holthausen:2012dk,Lam:2012ga,Lavoura:2014kwa,Byakti:2016rru,Jurciukonis:2016wrh} for many groups and analytically  for some groups in \cite{Hernandez:2012sk,Hernandez:2012ra}.

The main aim of this work is to present a simple and widely applicable general analytic discussion of the assumption $G_\nu=Z_2$, $G_l=Z_m$ for the type C and D subgroup of $SU(3)$\footnote{Earlier work  \cite{Yao:2015dwa} in this direction presented some analytic discussion of the possible mixing patterns in the type C, D and the group series $\Sigma(3 n^3)$.} as classified in \cite{Grimus:2010ak,Grimus:2011fk,Grimus:2013apa,Ludl:2011gn} and for all the known \cite{Ludl:2010bj} group series, namely $\Sigma(3 n^3)$, $T_n(m)$, $\Delta(3n^2,m)$, $\Delta(6 n^2,m)$ and $\Delta(6 n^2,j,k)$ of groups which are contained in $U(3)$ but not in $SU(3)$\footnote{ 
Such groups were earlier used to embedd the residual symmetries of a massless neutrino \cite{Joshipura:2013pga,Joshipura:2014pqa,King:2016pgv}.}.  The above stated assumption determines modulus of the elements in one of the columns of the leptonic mixing matrix. The analytic mixing formula derived here are used to show  that all possible predictions of this column in all the above groups can always be derived using some member of the the group $\Delta(6 n^2)$ making this  group series as optimal choice for the flavour symmetry $G_f$. This analytical discussion is supplemented with numerical analysis performed by scanning large number of known DSG of $U(3)$ and their viability is discussed in the context of the latest global fit to neutrino oscillation data. Results for some of the groups which are not described by the above series are discussed numerically.

We discuss general formalism of fixing the leptonic mixing matrix using the residual symmetries of neutrinos and charged leptons in the next section. Detailed analytic treatments of various groups are presented in section \ref{analytic} which is then followed by numerical investigations in section \ref{numerical}. We discuss in section \ref{additional} additional symmetries fixing the unknown observables which are not determined from $G_\nu=Z_2$ and summarize our results in section \ref{summary}.

\section{Formalism}
The leptonic mixing matrix can be determined from the given residual symmetries of neutrinos and charged leptons \cite{Lam:2007qc,Lam:2008rs,Lam:2008sh,Lam:2012ga,Lam:2011ag}. Consider three generations of massive Majorana neutrinos and the charged leptons both transforming as faithful irreducible representations of a discrete flavour group $G_f$. The $3 \times 3$ matrices $S_\nu$  and $T_l$ are elements of $G_f$ such that $S_\nu^2 = T_l^m = {\bf 1}$. Assume that the neutrino mass matrix $M_\nu$ and the charged lepton mass matrix $M_l$ are invariant under the action of symmetry such that 
\be \label{invariance}
S_\nu^T M_\nu S_\nu = M_\nu~~~~~{\rm and }~~~~~T_l^\dagger M_lM_l^\dagger T_l = M_l M_l^\dagger~. \ee
It is assumed that $S_\nu$ and $T_l$ do not commute in order to induce non-trivial mixing among the leptons. We further assume that all the eigenvalues of $T_l$ are distinct. The eigenvalues of $S_\nu$ are $\pm 1$.  We do not consider a trivial case corresponding to $S_\nu = -{\bf 1}$ as it does not provide any constraint on $M_\nu$. Therefore, two of the eigenvalues in $S_\nu$ are always degenerate and the other is unique. Let $V_\nu$ and $V_l$ be unitary matrices which diagonalizes $S_\nu$ and $T_l$
\be \label{vlvnu}
V_\nu^\dagger S_\nu V_\nu=d_\nu~,~~~~ V_l^\dagger T_lV_l=d_l~.\ee
Here $d_\nu$ and $d_l$ are diagonal matrices. Note that $V_\nu$ and $V_l$ are arbitrary upto a multiplication by a diagonal phase matrix from right hand side. Further, degeneracy of eigenvalues in $d_\nu$ imply that $V_\nu$ can also be multiplied by a unitary rotation in the space of the degenerate eigenvalues. The matrices $U_\nu$ and $U_l$ diagonalizing the leptonic mass matrices of eq. (\ref{invariance}) can therefore be written as
\be \label{unuul}
U_\nu=V_\nu P_\nu U_{ij}(\theta,\beta)~,~~ U_l=V_l P_l~,\ee
where $P_{l,\nu}$ are diagonal phase matrices. This leads to the leptonic mixing matrix $U_{\rm PMNS}$
\be \label{PMNS}
U_{\rm PMNS} = U_l^\dagger U_\nu~, \ee
The $U_{\rm PMNS}$ depends on the unknown phases in $P_\nu\equiv {\rm diag.}(e^{i\chi_1},e^{i\chi_2},e^{i\chi_3})$, the phase $\beta$ and angle $\theta$ contained in $U_{ij}$ where $ij$ refers to indices corresponding to degenerate eigenvalues in $d_\nu$. However modulus of one column of  $U_{\rm PMNS}$ is independent of these unknowns. If $c_\nu$ is an eigenvector of $S_\nu$ corresponding to the unique eigenvalue in $d_\nu$ then
\be \label{c0}
|c_0| = |V_l^\dagger c_\nu|~ \ee
represents the absolute values of the elements in a column of $U_{\rm PMNS}$ which is independent of the unknowns. In a special situation corresponding to $\beta=0$ in eq. (\ref{unuul}), one can also define further invariants which are independent of the mixing angle $\theta$. These are given by
\be \label{Aalpha}
A_\alpha={\rm Im}((U_{\rm PMNS})_{\alpha i}(U_{\rm PMNS})_{\alpha j}^*)~,\ee
where $\alpha=e,\mu,\tau$. These have been used earlier to derive useful conclusions in case of two degenerate neutrinos \cite{Hernandez:2013vya,Joshipura:2014qaa}. We shall use them  to obtain $\theta$ independent restrictions on the elements of $U_{\rm PMNS}$.  Although the absolute values of elements in $c_0$ are fixed, one can permute the elements within the column and rearrange them in order to comply with the observed $U_{\rm PMNS}$. Our aim in this paper is to derive possible predictions for $|c_0|$ using various  discrete DSG of $U(3)$ as $G_f$.

\section{An analytic study for fixed column prediction}
\label{analytic}
In this section, we obtain a universal formula for  a fixed column $|c_0|$ of $U_{\rm PMNS}$ applicable to DSG of $SU(3)$ categorized as  C and D and to a large number of DSG of $U(3)$ to be specified below. Apart from five exceptional groups, C and D types are the only  DSG of $SU(3)$  which have a 3 dimensional faithful irreducible representation. These are generated \cite{Grimus:2013apa} by the following matrices:
\beqa \label{generators}
E=\left(
\ba{ccc}
0&1&0\\
0&0&1\\
1&0&0\\
\ea \right)&,&~~~ B=-\left(
\ba{ccc}
1&0&0\\
0&0&1\\
0&1&0\\
\ea \right)~,\nonumber\\ \\
F=\left(
\ba{ccc}
\epsilon&0&0\\
0&\epsilon^k&0\\
0&0&\epsilon^{-k-1}\\
\ea \right)
&,&~~~
G=\left(
\ba{ccc}
1&0&0\\
0&\epsilon^{-r}&0\\
0&0&\epsilon^{r}\\ \ea \right)~.\eeqa
Here, $\epsilon=e^{2 \pi i/m}$, $r=m/n$ with $m$, $n$ and $r$ are nonzero and positive integers. The $k$ and $l_1$ are positive integers satisfying
$$1+k+k^2=l_1  r$$ with $k=0,1,...,r-1$. Type C groups are generated from the multiple products of $E$, $F$, $G$ with $m$, $n$, $k$ satisfying the above restrictions. Type D groups involve one more generator $B$ and also need additional restriction on $k$
$$2k+1=l_2 r$$
with $l_2$ an arbitrary positive integer. 

Our derivation is based on the observation that the  the elements of  groups generated by the multiple products of  the above generators can all be represented by six different textures defined  as ${\cal S} = \{R,S,T,U,V,W\}$ with elements in the set having the following form:
\beqa \label{textures}
W(\eta_1,\eta_2,\eta_3)=\left(\ba{ccc}
\et_1&0&0\\
0&\et_2&0\\
0&0&\et_3\\ \ea \right)&~,~~~~& V(\eta_1,\eta_2,\eta_3)=\left(\ba{ccc}
0&\et_1&0\\
0&0&\et_2\\
\et_3&0&0\\ \ea \right)~,\nonumber \\
S(\eta_1,\eta_2,\eta_3)=-\left(\ba{ccc}
\et_1&0&0\\
0&0&\et_2\\
0&\et_3&0\\ \ea \right)&~,~~~~&R(\eta_1,\eta_2,\eta_3)=\left(\ba{ccc}
0&0&\eta_1\\
\eta_2&0&0\\
0&\eta_3&0\\ \ea \right)~,\nonumber \\
T(\eta_1,\eta_2,\eta_3)=-\left(\ba{ccc}
0&0&\eta_1\\
0&\eta_2&0\\
\eta_3&0&0\\ \ea \right)&~,~~~~&U(\eta_1,\eta_2,\eta_3)=-\left(\ba{ccc}
0&\eta_1&0\\
\eta_2&0&0\\
0&0&\eta_3\\ \ea \right) ~.\eeqa%
These textures have the properties that only one entry in a given column and row is non-zero and it is labeled by $\eta_i$ which is an arbitrary root of unity. Only the exact expressions of $\eta_i$ vary for different groups and for different elements within the groups. The above textures do not just describe the elements of the groups of type C and D but have larger validity and can describe a large number of DSG of $U(3)$ and not just $SU(3)$.  This follows from the expressions of generators of the DSG of $U(3)$ given by \cite{Ludl:2010bj}. These generators have the same textures as above with replacement of $\eta_i$ by powers of the $n^{th}$ roots of unity. Combinations of these generators with in general different integer $n$ also lead to elements having the same textures as above \cite{Joshipura:2014pqa}. All the groups which can be obtained from the above set of generators and having order $<512$ are listed in  \cite{Ludl:2010bj}. In addition, elements of the five infinite series of the DSG of $U(3)$  
identified in  \cite{Ludl:2010bj} can also be parameterized as in eq. (\ref{textures}). Thus the above textures describe a large bulk of DSG of $SU(3)$ and $U(3)$. For example, total 55 out of 59 DSG of $SU(3)$ and 70 out of 75 DSG of $U(3)$ (which are not DSG of $SU(3)$) listed in \cite{Ludl:2010bj} with order less than 512 belong to this category. 

We now derive analytic expressions for the possible patterns of a column in $U_{\rm PMNS}$ if $S_\nu$ and $T_l$ have textures specified in ${\cal S}$. Since $S_\nu^2={\bf 1}$, it forms a $Z_2$ subgroup of a given flavour group $G_f$ which is a DSG of $SU(3)$ or $U(3)$. The elements $V$ and $R$ in ${\cal S}$ do not form $Z_2$ subgroups as their eigenvalues are all different and are given by det.$(V)(1,\omega,\omega^2)$ and det.$(R)(1,\omega,\omega^2)$ respectively, with $\omega=e^{2 \pi i/3}$. Thus $S_\nu$ can be any of $W$ with the eigenvalues $(-1,-1,1)$ or any of $S$, $T$ and $U$ with specific values of $\eta_i$. For example,
\be
\label{s}
S(1,\eta_\nu^*,\eta_\nu)=-\left(
\ba{ccc}
1&0&0\\
0&0&\eta_\nu^*\\
0&\eta_\nu&0\\
\ea \right)~\ee
and equivalent $T$ and $U$ obtained by permuting the above $S$ have eigenvalues $(-1,-1,1)$. The symmetry of the charged leptons $T_l$ can have any structure within ${\cal S}$ with all the eigenvalues  being distinct and $[S_\nu, T_l]  \neq 0$.

As discussed in the previous section, the lepton mixing matrix depends on the matrices diagonalizing $S_\nu$ and $T_l$. If $V_G$ is a unitary matrix that diagonalizes a given element $G \in {\cal S}$ such that $V_G^\dagger G V_G=G_{\rm diag.}$ then different $V_G$ are given by \cite{Joshipura:2014pqa}
\beqa \label{ug}
V_W={\bf 1}&,&~~~~ V_V={\rm Diag. }(1,\eta_1^*\Delta,\eta_2^*\eta_1^*\Delta^2) U_\omega,\nonumber\\
V_S=\frac{1}{\sqrt{2}}\left( 
\ba{ccc}
\sqrt{2}&0&0\\
0&1&1\\
0&\eta_2^*\lambda_{23}&-\eta_2^*\lambda_{23}\\
\ea\right)&,&~~~~ V_R={\rm Diag. }(1,\eta_1^*\eta_3^*\Delta^2,\eta_1^*\Delta) U_\omega,\nonumber\\
V_T=\frac{1}{\sqrt{2}}\left( 
\ba{ccc}
1&0&1\\
0&\sqrt{2}&0\\
\eta_1^*\lambda_{13}&0&-\eta_1^*\lambda_{13}\\
\ea \right)&,&~~~~ V_U=\frac{1}{\sqrt{2}}\left( 
\ba{ccc}
1&1&0\\
\eta_1^*\lambda_{12}&-\eta_1^*\lambda_{12}&0\\
0&0&\sqrt{2}
\ea\right).
\eeqa
Here $\lambda_{ij}=\sqrt{\eta_i\eta_j}$, $\Delta=(\eta_1\eta_2\eta_3)^{1/3}$ and $U_\omega$ is defined as 
\be \label{uw}
U_\omega=\frac{1}{\sqrt{3}}\left( \ba{ccc}
1&1&1\\
1&\omega & \omega^2\\
1&\omega^2 & \omega\\
\ea \right).\ee

It is possible to work out all possible patterns of a fixed column using the above diagonalizing matrices and eq. (\ref{c0}). The column $c_\nu$ in eq. (\ref{c0}) refers to an eigenvector corresponding to a unique eigenvalue in $S_\nu$. The following two possibilities exist for $c_\nu$ corresponding to $S_\nu =W(-1,-1,1)$ or $S_\nu \in \{ S, T, U \}$ with the particular structure as specified in eq. (\ref{s}):
\be \label{cnu}
c_\nu = (0,0,1)^T~~~{\rm or}~~~c_\nu = \left(0,  \frac{1}{\sqrt{2}},- \frac{\eta_\nu}{\sqrt{2}} \right)^T, \ee
or with permutations of the elements within a given $c_\nu$.  The different possible choices of $T_l$ then lead to the following structures for fixed column $|c_0|$ according to eq. (\ref{c0}). 
\begin{enumerate}[(1)]
\item For $S_\nu = W(-1,-1,1)$ and $T_l \in \{ R, V\}$, 
\be \label{predict1}
|c_0|^2 =\left( \frac{1}{3},\frac{1}{3},\frac{1}{3}\right)^T\ee
\item For $S_\nu = W(-1,-1,1)$ and $T_l \in \{ S, T, U\}$ or $S_\nu \in \{ S, T, U\}$ and $T_l = W(\eta_1,\eta_2,\eta_3)$, 
\be \label{predict2}
|c_0|^2 = \left( 0,\frac{1}{2},\frac{1}{2}\right)^T~\ee
or its permutations.
\item For $S_\nu \in \{ S, T, U\}$ and $T_l = V(\eta_1,\eta_2,\eta_3)$, 
\be \label{predict3}
|c_0|^2 = \frac{1}{6} \left( |1 -\eta_\nu \eta_2\Delta^*|^2,|1 -\omega\eta_\nu \eta_2\Delta^*|^2,|1 -\omega^2\eta_\nu \eta_2\Delta^*|^2 \right)^T~\ee
or its permutations. The choice $T_l = R(\eta_1,\eta_2,\eta_3)$ leads to a similar structure with $\eta_2$  replaced by $\eta_3$.
\item For $S_\nu \in \{ S, T, U\}$ and $T_l \in \{ S, T, U\}$ but with textures different than $S_\nu$,
\be \label{predict4}
|c_0|^2 = \left( \frac{1}{2},\frac{1}{4},\frac{1}{4}\right)^T~\ee
or its permutations.
\item For $S_\nu \in \{ S, T, U\}$ and $T_l \in \{ S, T, U\}$ but with textures same as $S_\nu$,
\be\label{predict5}
|c_0|^2=\frac{1}{4}\left(0, |1+\sqrt{\eta_2^*\eta_3^*} \eta_\nu |^2,|1-\sqrt{\eta_2^*\eta_3^*} \eta_\nu|^2 \right)^T\ee
or its permutations.
\end{enumerate}
The above five equations exhaust all the predictions for one of the columns of $U_{\rm PMNS}$ for all the groups under considerations. Three of the predictions are universal in the sense that they are independent of the choice of $\eta_{1,2,3}$ and $\eta_\nu$ in all the formula above.  Among them eq. (\ref{predict1}), namely the tri-maximal column, provides a viable prediction for the second column of $U_{\rm PMNS}$. The bi-maximal column in eq. (\ref{predict1}) at best can be identified with the third column of $U_{\rm PMNS}$ with large correction required through breaking of the residual symmetries. The prediction in eq. (\ref{predict5}) depends on the free parameters but one of the entries of the column is predicted to be zero and thus it also requires large correction. The solution in eq. (\ref{predict3}) offers several possibilities which can lead to correct leading order predictions for the columns of $U_{\rm PMNS}$ matrix. The complex numbers $\eta_i$ are some roots of unity. They do not represent the 
same root nor powers of some specific roots of unity in general. Their exact expressions are specified by the properties of the groups and we discuss various cases in the next subsections. We also show that eq. (\ref{predict3}) can fit either the first or third column of PMNS matrix at the leading order. Therefore, tri-maximal is the only solution offered by the groups under considerations which can be consistently identified with the second column of the $U_{\rm PMNS}$ matrix.

\subsection{Mixing angle predictions for the DSG of $SU(3)$}
Among the DSG of $SU(3)$, the groups of type C and D are generated \cite{Grimus:2013apa} using two or more elements from the textures given in eq. (\ref{textures}). We discuss below both these group series.

\subsubsection{Type C groups}
The type C groups defined earlier are isomorphic to $(Z_m\times Z_n)\rtimes Z_3$ where $Z_m$, $Z_n$ are respectively generated by $F$, $G$ in eq. (\ref{generators}) and $Z_3$ by $E$. This isomorphism can be used to find all elements of the C type groups by using the result that any element $s\in S=G\rtimes H $ can be uniquely written as $s=g h $ in terms of $g\in G$ and $h\in H$. Thus all the elements $\in C$ are written as
$$ F^p G^q E^a$$
with $p=0,1,...,m-1$; $q=0,1,...,n-1$ and $a=0,1,2$. It is then seen that all the elements fall in three
textures labeled as $W$, $R$ and $V$  in eq. (\ref{textures}) with the identification
\be \label{identification}
\eta_1=\epsilon^p,~~\eta_2=\epsilon^{kp-r q},~~\eta_3=\epsilon^{-p(k+1)+r q}~.\ee

Since $R$ and $V$ do not form $Z_2$ subgroups only possible candidates for $Z_2$ subgroups are the diagonal generators $W$ when $m$ and $n$ are even. The resulting mixing pattern is always tri-maximal as in eq. (\ref{predict1}). This leads to two predictions:
\be \label{tri-maximal}
c_{13}^2s_{12}^2=\frac{1}{3}~~~{\rm and}~~~\cos\delta=
\frac{\left(c_{23}^2-s_{23}^2\right) \left(c_{12}^2-s_{12}^2
 s_{13}^2\right)}{4 c_{12} c_{23} s_{12} s_{13} s_{23}}~,\ee
where $s_{ij} =\sin\theta_{ij}$ and $c_{ij}=\cos\theta_{ij}$. The first leads to a solar angle quite close to its global fit value \cite{Esteban:2016qun}. The second prediction correlates the Dirac CP phase with the values of  $s_{23}^2$ and $s_{13}^2$. The sign of $\cos\delta$ is strongly correlated to the quadrant of the atmospheric mixing angle and one gets positive (negative) values for $\cos\delta$ for $\theta_{23}$  smaller (greater) than $45^\circ$.  It is to be noted that  T2K results prefer a  negative $\cos\delta$ \cite{Abe:2013hdq} and the global best fit value of $\theta_{23}$ lies in the second quadrant according to \cite{Esteban:2016qun}. Both these results are consistent with eq. (\ref{tri-maximal}). However either sign  of $\cos\delta$ and both quadrants for $\theta_{23}$
are allowed at 3$\sigma$.

Let us mention possible groups obtained for the lowest values of $r$. $r=1$  leads to $\Delta(3 n^2)$ series and the lowest member with $Z_2$ subgroup is $A_4$. For $r=3$, either one gets the series 
$Z_3\times \Delta(3 n^2)$ or the series $(Z_{9n}\times Z_{3 n})\rtimes Z_3$ and the lowest member of the latter containing $Z_2$ subgroup is an order 324 group. Other possible group series of type C are listed in \cite{Grimus:2013apa}.

\subsubsection{Type D groups}
These groups are generated by $F$, $G$, $E$, $B$ of eq. (\ref{generators}) and are isomorphic to $(Z_m\times Z_n)\rtimes S_3$. $S_3$ is now generated by $E$ and $B$. As before, we can explicitly write the expressions for all the elements of the D type groups, $(Z_m\times Z_n) \rtimes S_3$ in terms of their generators. To do this, we note that $S_3$ can be written as $Z_3\rtimes Z_2$. Hence all its six elements can be written as $E^aB^l$, where $a=0,1,2$ and $l=0,1$. Then the semi-direct product structure of D implies that all its elements can be uniquely written as
\be \label{d-elements}
F^p G^q E^a B^l~.\ee
One sees from the explicit forms of generators that all these elements are given by the six textures
in eq. (\ref{textures}). But now  $\eta_{1,2,3}$ are given in terms of $p$, $q$ by eq. (\ref{identification}). All possible mixing patterns within the type D groups are generated by
substituting these values of $\eta_1$, $\eta_2$, $\eta_3$ in eqs. (\ref{predict1}-\ref{predict5}) and varying $p$, $q$ for a given group, {\it i.e.} for given values of $k$, $m$ and $n$. Due to restrictions on the values of $r$ and $k$, essentially two type-D group series exist \cite{Grimus:2013apa}. They are obtained with $k=0$, $r=1$ and $k=1$, $r=3$. These are respectively isomorphic to $\Delta(6n^2)\equiv (Z_n \times Z_n)\rtimes S_3$ and $D^1_{9n,3n}\equiv (Z_{9n}\times Z_{3n}) \rtimes S_3$. It is possible to write down simple expressions for the predicted mixing for both these series and we discuss them below.

Let us first consider the case of $\Delta(6n^2)$.
\begin{itemize}
 \item Since $\Delta(3n^2)$ are subgroups of $\Delta(6 n^2)$, the universal prediction given by eq.  (\ref{predict1}) is obtained in all $\Delta(6n^2)$ groups with even $n$ as in the case of $\Delta(3 n^2)$.
 \item  The $Z_2$ subgroups generated by $B$, $EBE^{-1}$ and $E^2BE^{-2}$ exist in all the $\Delta(6n^2)$ groups and consequently the predictions given in eqs. (\ref{predict2},\ref{predict4}) follow in all $\Delta(6n^2)$ groups. In addition, groups with $n>2$ also allow values of $\eta_i$, $\eta_\nu$ other than $\pm 1$ and lead to predictions as given by eq.  (\ref{predict5}). This allows possibility of predicting non-maximal $\theta_{23}$ but the $\theta_{13}$ remains zero in all these cases.  
\item The non-universal prediction as given by eq. (\ref{predict3}) can lead to non-trivial expressions for the columns of $U_{\rm PMNS}$. It is obtained for the choice $S_\nu=S(1,\eta^{-q'},\eta^{q'})$ and $T_l=V(\eta_1,\eta_2,\eta_3)$. Here $\eta_{1,2,3}$ are given by eq. (\ref{identification}) with  $k=0$, $r=1$ and $m=n$. Substituting these in  eq. (\ref{predict3}), we obtain
\be \label{6nsquarepredict3}
|c_0|^2 = \frac{1}{6}\left(|1 -\epsilon^{q'-q}|^2,~|1 -\omega \epsilon^{q'-q}|^2,~|1 -\omega^2 \epsilon^{q'-q}|^2\right)~,\ee
where $q,q'=0,1,..., n-1$.
 \end{itemize}

\begin{figure}[!ht]
\centering
\includegraphics[width=0.75 \textwidth]{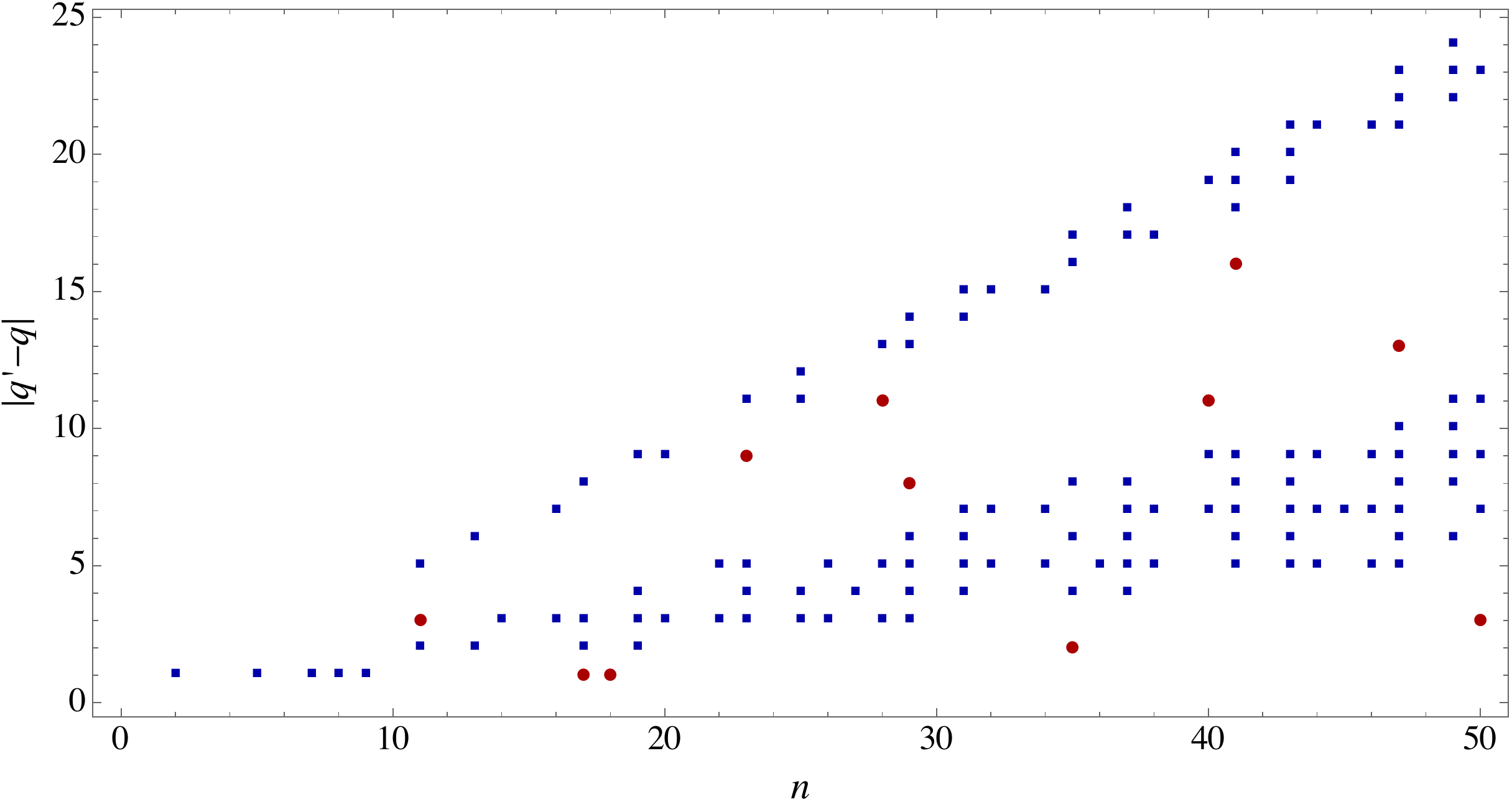}
\caption{The values of $n \le 50$ and $|q'-q|$ in eq. (\ref{6nsquarepredict3}) leading to the viable columns of leptonic mixing matrix. The blue squares (red dots) indicate that the corresponding prediction is consistent with the first (third) column of $U_{\rm PMNS}$ matrix within $3\sigma$. Each point represents a unique solution obtained by the smallest possible values of $n$ and $|q'-q|$.}
\label{fig1}
\end{figure}

The mixing patterns implied by the above equation are obtained by varying $n$ and correspondingly $q$ and $q'$. Not all of these give physically different predictions. Replacement  $\epsilon \leftrightarrow \epsilon^*$ permutes the last two entries in eq. (\ref{6nsquarepredict3}). Whenever $n$ is divisible by 3 then $\epsilon^{q-q'}$, $\omega \epsilon^{q-q'}$, $\omega^2 \epsilon^{q-q'}$ are all $n^{\rm th}$ roots of unity. In such case, {\it i.e.} $n=3p$, one gets $p/2$ ($(p-1)/2$) number of non-trivial solutions with all non-zero entries for $p$ even (odd). If $n$ is not divisible by 3, then eq.  (\ref{6nsquarepredict3}) leads to $n/2$ ($(n-1)/2$) non-trivial solutions with all non-zero entries for $n$ even (odd). These observations agree with the results of numerical calculations given in the next section. It is to be noted that eq. (\ref{6nsquarepredict3}) does not contain the tri-maximal
solution $|c_0|^2=(1/3,1/3,1/3)$ for any value of $|q-q'|$. It has to come from  eq. (\ref{predict1}) which requires $n$ to be even. It is straightforward to work out all solutions of eq. (\ref{6nsquarepredict3}) for different $n$. We show in Fig. \ref{fig1} the values of $n$ and $|q'-q|$ leading to the predictions for a column of lepton mixing matrix which is consistent with a column of the experimentally obtained leptonic mixing matrix, namely $U^{\rm exp}_{\rm PMNS}$. 
The recent global fit ``NuFIT 3.0" of neutrino oscillation data by NuFIT collaboration \cite{Esteban:2016qun} (for the latest global fit results, see NuFIT 3.0 (2016), \url{http://www.nu-fit.org}) allows the following $3\sigma$ ranges in the elements of $U_{\rm PMNS}$.
\be \label{expUPMNS}
|U^{\rm exp}_{\rm PMNS}| = \left( \ba{ccc}
0.800 \to 0.844 & 0.515 \to 0.581 & 0.139 \to 0.155\\
0.229 \to 0.516 & 0.438 \to 0.699 & 0.614 \to 0.790\\
0.249 \to 0.528 & 0.462 \to 0.715 & 0.595 \to 0.776\\
\ea \right)
\ee

We find that eq. (\ref{6nsquarepredict3}) can predict either the first or the  third column of $U_{\rm PMNS}$. The simplest example corresponds to $n=2$. This leads to only one non-trivial solution given by $|c_0|^2=(2/3,1/6,1/6)^T$ corresponding to the first column of the famous tri-bimaximal mixing. The smallest group in $\Delta(6n^2)$ series which reproduces a viable third column of $U_{\rm PMNS}$ corresponds to $n=11$. It gives a fixed column $|c_0|^2=(0.0239,0.3808,0.5954)^T$ leading to the values $\sin^2\theta_{13}=0.0239$ and $\sin^2\theta_{23}= 0.39$ or $0.61$. We discuss this case in more detail in the next section. It is straightforward to show that eq. (\ref{6nsquarepredict3}) cannot reproduce the second column of $|U^{\rm exp}_{\rm PMNS}|$ for any value of $n$ and $(q'-q)$. The differences between two of the elements of $|c_0|^2$, defined as $\Delta_{ij} = |c_{0i}|^2 - |c_{0j}|^2$, are given by
\beqa \label{diff}
\Delta_{12} &=& - \frac{1}{2} \left( \cos \xi + \frac{1}{\sqrt{3}} \sin \xi \right), \nonumber \\
\Delta_{13} &=& - \frac{1}{2} \left( \cos \xi - \frac{1}{\sqrt{3}} \sin \xi \right), \nonumber \\
\Delta_{23} &=& \frac{1}{\sqrt{3}}\sin\xi~,\eeqa
where $\xi = 2\pi (q'-q)/n$. The $3\sigma$ ranges allowed in $|U^{\rm exp}_{\rm PMNS}|$ necessarily require all the three $|\Delta_{ij}|< 0.277$ if the predicted column is identified with the second column of $U_{\rm PMNS}$. These three conditions cannot be simultaneously satisfied by any value of $\xi$ in eq. (\ref{diff}). Therefore  eq. (\ref{6nsquarepredict3}), and more generally eq. (\ref{predict3}), does not lead to any prediction for the second column of $U_{\rm PMNS}$ matrix which is consistent with the current global fit results.
 
The remaining groups of type D are given by the series $D^1_{9n',3n'}$  isomorphic to $(Z_{9n'}\times Z_{3n'})\rtimes S_3$. We now show that mixing angle predictions given in eqs. (\ref{predict1}-\ref{predict5}) in these groups coincide with the corresponding ones obtained in the case of $\Delta(6n^2)$ with $n=9n'$. All the elements of the groups are given by eq.  (\ref{d-elements}). The $\eta_i$ for these groups are given by eq.  (\ref{identification}) with substitution $k=1$, $r=3$ and $m=9 n'$, $n=3 n'$. The symmetry of neutrinos corresponding to eq.  (\ref{predict5}) is $S_\nu=S(1,\eta^{-q'},\eta^{q'})$. Substituting these values in eq.  (\ref{predict5}), we get
\be \label{predic5-D1}
|c_0|^2 = \frac{1}{6}\left( |1-\epsilon^{3(q-q')-p}|^2 , |1-\omega \epsilon^{3(q-q')-p}|^2, |1-\omega^2\epsilon^{3(q-q')-p}|^2\right)~,\ee
where $p=0,1,...,9n'-1$ and $q,q'=0,1,...,3n'-1$. It is noted that $3(q-q')-p = 0,1,...,9n'-1$. Therefore, the above expression of the predicted column coincides with the prediction eq.  (\ref{6nsquarepredict3}) obtained in the case of  $\Delta(6n^2)$ series with $n=9 n'$. A similar argument also holds for 
the prediction given in eq. (\ref{predict5}). The smallest group in the series  $D^1_{9n',3n'}$ is order 162 group obtained for $n'=1$.
It predicts the same column $|c_0|^2$ as the group $\Delta(486)$ and by the counting argument given above there is only one nontrivial prediction with all non-zero entries. This is  given in the Table \ref{su3D-result}. The next higher group with $n'=2$ is an order $648$ group and it has three independent non-trivial predictions which can also be obtained from $\Delta (6\times 18^2)$.  As can be seen from Fig. \ref{fig1}, this group will give only a third column consistent with the $3\sigma$ range of experimental data. Similar arguments holds for other predictions as well and thus group series $D^1_{9n',3 n'}$ are equivalent to the group series $\Delta(6\times (9n')^2)$. This is expected since all the D types of groups  are  shown to be \cite{Zwicky:2009vt} subgroups of $\Delta(6n^2)$ groups for some $n$.

\subsection{Mixing in DSG of $U(3)$}
We now analyse the DSG of $U(3)$ which are not subgroups of $SU(3)$. Ludl \cite{Ludl:2010bj} has identified several group series belonging to this category. Elements in
all of these series have textures  given in eq. (\ref{textures}). Hence the mixing formulas eq. (\ref{predict1}-\ref{predict5}) are applicable to them. Using these,  we now argue that none of the groups in these  series of groups lead to  any new mixing patterns compared to the ones already found with   DSG of $SU(3)$. The same result is also found to be numerically true for all the groups of order less than 512 and not just for the groups belonging to the said series as we discuss in the next section.  This can  be shown in a more general way using the following redefinition of the residual symmetries of the charged leptons and neutrinos\footnote{We thank referee for pointing out this general result.}. Let $S_\nu$ be a generator of $G_\nu=Z_2$ and $T_l$  a generator of $G_l=Z_m,~m \ge 3$. Both $G_\nu$ and $G_l$ are subgroups of some $G_f$ which is DSG of $U(3)$ and not of $SU(3)$.  One can define
\be \label{defineprime}
S_\nu^\prime = ({\rm det.}(S_\nu))~S_\nu~~~~~{\rm and}~~~~~T_l^\prime = ({\rm det.}(T_l))^{-\frac{1}{3}}~T_l~, \ee
where ${\rm det.}(S_\nu)$ is $\pm 1$. 
det.($S_\nu^\prime$) = det.($T_l^\prime)$ are now $+1$ and $G_f^\prime$
containing them is a DSG of $SU(3)$. They lead to the same invariance as in
eq. (\ref{invariance}). Both  $S_\nu^\prime$ and $T_l^\prime$
are respectively diagonalized by the same matrices $U_\nu$ and $U_l$ which
diagonalize $S_\nu$ and $T_l$ resulting in the same
prediction for a column of $U_{\rm PMNS}$  as obtained from $G_f$.
Therefore, all the predictions of $G_f$ belonging to $U(3)$ are always
obtained from some $G_f^\prime$ within $SU(3)$. Above mentioned argument also remains true if the residual
symmetry of the neutrino mass matrix is a  Klein group $Z_2\times Z_2$
generated by $S_{1\nu}$ and $S_{2\nu}$ contained in some DSG of $U(3)$. One
can define new generators $S'_{1\nu},S'_{2\nu}$ and $T_l'$ as in
eq. (\ref{defineprime}) having  determinant $+1$. These generators would be
contained in some DSG of $SU(3)$ and lead to the same
mixing pattern as given by a DSG of $U(3)$ containing 
$S_{1\nu},S_{2\nu},T_l$.

Although the  DSG of $U(3)$ do not offer any new prediction than those obtained from the DSG of $SU(3)$ as shown above, it is still interesting from the practical point of view to know which of the five sets of predictions given in eqs. (\ref{predict1}-\ref{predict5}) can be obtained from a given DSG of $U(3)$. We continue our analytical treatment in the case of DSG of $U(3)$ in order to explore the connection between the given DSG of $U(3)$ and their $SU(3)$ counterparts which offer the same sets of predictions. Here, we specifically show that 
\begin{itemize}
 \item All the groups in series $\Delta(3 n^2,m)$ give the same pattern of column as $\Delta(3 n^2)$, namely only tri-maximal for even $n$.
 \item The group series $\Delta(6 n^2,m)$ does not contain any $Z_2$ subgroups (except the trivial one corresponding to $S_\nu = -{\bf 1}$) for odd $n$. For even $n$, it gives only the tri-maximal and the bi-maximal columns.
 \item The series  $\Delta'(6 n^2,j,k)$ contain $\Delta(6 n^2)$ as subgroup only  if $k=1$. In that case, one gets the same mixing pattern as $\Delta(6 n^2)$. The groups with $k>1$, either do not contain any  non-trivial $Z_2$ subgroups or they are diagonal when they exist. In the latter case one gets either the bi-maximal or  the tri-maximal mixing  pattern.
 \end{itemize}

%

\subsubsection{The group series $\Delta(3 n^2,m)$}
These groups are isomorphic to $(Z_n\times Z_n)\rtimes Z_{3^m}$. Here $n$ is not an integer multiple of 3 and $m$ is an integer $>1$. Its generators are 
\be
\delta_m E,~F(n,0,1)=W(1,\eta,\eta^*) \ee
with $\delta_m=e^{2 \pi i/3^m}$ and $\eta=e^{2 \pi i/n}$. Semi-direct structure implies that all the group elements can be written as
\be 
X^jY^k(\delta_m E)^l \ee
where $X=W(1,\eta,\eta^*)$, $Y=EXE^{-1}=W(\eta^*,\eta,1)$ and $j,k=0,1,...,n-1$ while  $l=1,2,...,3^m$. Using the fact that
\beqa
(\delta_m E)^l&=&\delta_m^l E~~~{\rm for}~~l=1,4,7,...~\nonumber\\
(\delta_m E)^l&=&\delta_m^l E^2~~~{\rm for}~~l=2,5,8,...\nonumber\\
(\delta_m E)^l&=&\delta_m^l ~~~{\rm for}~~l=3,6,9,...\nonumber~ \eeqa
one sees that all the $3^m n^2$  elements of $\Delta( 3 n^2,m)$ groups have the same texture as
in $\Delta(3 n^2)$ apart from an over all multiplication by $\delta_m^l$. As a consequence, only $Z_2$ groups are diagonal if $n$ is even and no non-trivial $Z_2$ subgroup exists if $n$ is odd. Only mixing pattern in the former case is tri-maximal as  was shown in case of $\Delta(3n^2)$ group series.

\subsubsection{The group series $\Delta(6 n^2,m)$}
These groups are isomorphic to $\Delta(3 n^2)\rtimes Z_{2^m}$ where $n$ is not an integer multiple of 3  and $m$ is an integer $>1$. The elements of these groups are generated by 
\be
E,~F(n,0,1),~\beta_m T \ee
with $\beta_m=e^{2 \pi i/2^m}$ and 
\be \label{t}
T =T(0,0,0)=-\left(\ba{ccc}
0&0&1\\
0&1&0\\
1&0&0\ea\right)~.\ee
Because of the semi-direct product structure, the elements can be written as 
\be
g=X^jY^kE^l (\beta_m T)^r,~\ee
with $X,Y$ as defined in the previous case, $j,k=0,1,...,n-1$; $l=0,1,2$ and
 $r=1,1,...,2^m$. It is seen that $(\beta_m T)^r= \beta_m^r T$ if $r$ is
odd and $(\beta_m T)^r= \beta_m^r$ if $r$ is even. As a result of this, the
elements can be divided into two groups
\beqa \label{elements-6n2m}
g&=& \beta_m^{2 k} (W,R,V)~,\nonumber \\
&=& \beta_m^{2 k-1} (WB,RB,VB)~,\eeqa
Here, $k=1,2,...,2^{m-1}$ and $(W,R,V,WB,RB,VB) $ are six textures
characterizing elements $X^jY^kE^lT^q$, $(q=0,1)$ of the $\Delta(6 n^2)$
groups. The $\Delta(6 n^2)$ group is however not contained in $\Delta(6
n^2,m)$ due to the factor $\beta^{2 k-1}_m$  (which is $\neq \pm
1$ for any $k$) multiplying the second line of the above equation. As a
result the only  non-trivial $Z_2$ subgroups are diagonal and
are obtained from the first line in eq.  (\ref{elements-6n2m}) 
 corresponding to even $r$. These are given by $W=W(1,-1,-1)$
and its permutations for $r=2k=2^m$ and by $-1,-W(1,-1,-1)$ and permutations
for $r=2k=2^{m-1}$. The $r$ is even in both the cases since $m> 1$.
 With any of these as $S_\nu$ and
$T_l =\beta_m^{2 k}(R,V)$ one gets the tri-maximal pattern. If $T_l$ is
chosen as any of the elements in the second line in eq. 
(\ref{elements-6n2m}), one gets bi-maximal pattern. These are thus the only
patterns possible with this group series. This is true for the series
$S_4(m)$ also which are special cases of $\Delta(6 n^2,m)$ with $n=2$.

\subsubsection{The group series $\Delta'(6 n^2,j,k)$}
These groups are generated by
\be
E,~X,~\beta_{j,k} T  \ee
Here $X$ is as defined earlier, $n$ is a multiple of 3 and $\beta_{j,k}=e^{\frac{2\pi i}{3^j 2^k}}$. $T$ is the standard $Z_2$ generator defined in eq. (\ref{t}). This group does not have  a semi-direct product structure as in the previous cases but still it is possible to write all its elements in the form
\be \label{jkelements}
g=X^mY^pE^q (\beta_{j,k} T)^r\ee
with $m=0,1...,n-1$; $p=0,1,...,\frac{n}{3}-1$; $q=0,1,2$ and $r=0,1,...,3^j
2^k-1$. This group therefore has $n^2 3^j 2^k$ elements. This is seen by
noting that the multiple products of $X$ and $E$ among themselves generate a
set of elements $X^s Y^t E^q$ with $s$, $t$ running over full range
$0,1,...,n-1$. Powers of $\beta_{j,k} T$ occurring anywhere between them can
be transferred to the end using $(\beta_{j,k} T) E(\beta_{j,k} T)^{-1}=E^2$ 
and $(\beta_{j,k} T)X(\beta_{j,k} T)^{-1}=Y$ etc.  Furthermore, $X$ and $Y$ 
are seen to satisfy
$$Y^{\frac{n}{3}}=X^{\frac{n}{3}}\omega^2 {\bf 1}=X^{\frac{n}{3}}(\beta_{j,k} T)^{3^{(j-1)}2^{k+1}}$$
Thus $Y^a$ with $a\geq n/3$ can be written in terms of powers of  $X$ and $ (\beta_{j,k} T)$ with $Y^b$, $b<n/3$. This allows one to restrict the range of $p$ in eq. (\ref{jkelements}) as given there.

The elements in eq.  (\ref{jkelements}) can be divided into two groups as before
\beqa \label{groups}
g&=& \beta_{j,k}^r (W,R,V)~~~{\rm for~even}~r,\nonumber\\
&=&\beta_{j,k}^r (WT,RT,VT) \sim \beta_{j,k}^r (T,U,S)~~~{\rm for~odd}~r~.\eeqa
We have used the fact that the $(WT,RT,VT)$ have the same structure as $(T,U,S)$ respectively.
It follows that
\begin{itemize}
\item If $k=1$ then an odd power of $\beta_{j,k}$ namely $\beta_{j,k}^{3^j}$ equals $-1$. In that case 
$\Delta'(6 n^2,j,1)$ contains $(W,R,V,-S,-T,-U)$ which form a $\Delta(6 n^2)$ subgroup. This is not the case for $k>1$ and $\Delta(6 n^2)$ is not a subgroup of $\Delta'(6 n^2,j,k)$.  The group $\Delta(3 n^2)$ is always a subgroup of $\Delta'(6 n^2,j,k)$ obtained when $r=0$.
\item All the elements in eq.  (\ref{groups}) have the same six textures as in eq. (\ref{textures}) with $\eta_i=\eta^a \beta_{j,k}^r$, $\eta^a$ being some power of the $n^{\rm th}$ root of unity. Mixing predictions for the group are given by substituting these $\eta_i$ in eqs. (\ref{predict3}) and eq. (\ref{predict5}) when $\Delta(6 n^2)$ is a subgroup, {\it  i.e.} for $k=1$. The overall factor $ \beta_{j,k}^r$ in $\eta_i$ is seen to cancel in these equations and the mixing angle predictions coincide in this case with the ones obtained for $\Delta(6 n^2)$ even though one has a larger set of $T_l$ to choose from. This can be explicitly seen from the numerical tables to be presented where one sees that $\Delta(54)$ and $\Delta(54,3^j,1)$ give identical mixing pattern namely bi-maximal. A slightly non-trivial example is the group $\Delta(6 \times 6^2,2,1)$ which is an order 648 group. All the solutions obtained for this group numerically coincide with the prediction of the $\Delta(6 \times 6^2)=\Delta(216)$ 
group displayed in the Table \ref{su3D-result}.
\item For $k>1$ and even $n$, the $Z_2$ subgroups are diagonal. The choice
$T_l=\beta_{j,k}^r (R,V)$ then lead to the tri-maximal pattern. The choice $T_l=\beta_{j,k}^r(T,U,S)$ lead to the bi-maximal mixing. These two provide the only possible mixing patterns within $\Delta'(6 n^2,j,k)$ with $k>1$.
\end{itemize}

Continuing similar argument, one can also show that   group series $\Sigma(3 n^3)$
 first studied in \cite{Ishimori:2010au} can only lead to the tri-maximal mixing pattern.
 The other remaining series identified by Ludl \cite{Ludl:2010bj} namely
$T_n(m)$, does not contain any $Z_2$ subgroup and cannot therefore contain
residual symmetry $S_\nu$.
\section{Numerical analysis}
\label{numerical}
We numerically analyze all the DSG of $U(3)$ having three dimensional IR and order $<512$ and derive their prediction for column in leptonic mixing matrix. These groups are listed in \cite{Ludl:2010bj} along with their generators. Using the given generators, we numerically generate all the elements for a given group $G_f$. We then identify its $Z_2$ elements as the symmetry $S_\nu$ of neutrinos and find an eigenvector $c_\nu$ corresponding to the unique eigenvalue in $S_\nu$. The $T_l$ are chosen such that $T_l^m=1$, $m \ge 3$, $[T_l,S_\nu] \neq 0$ and all the eigenvalues of $T_l$ are distinct. We then find a unitary matrix $V_l$ which diagonalizes $T_l$ and find a column $|c_0|$ using eq.  (\ref{c0}). The column $|c_0|$ or its permutation predicts a column in $U_{\rm PMNS}$ matrix. 

We first perform a scan over DSG of $SU(3)$.  Apart from the groups of type C and D which are already discussed analytically in the previous section, it contains six exceptional groups: $A_5$, $PSL(2,7)$, $\Sigma(36 \phi)$, $\Sigma(72 \phi)$, $\Sigma(216 \phi)$ and $\Sigma(360 \phi)$. The last two groups have order $> 512$. The results of scan are listed in Table \ref{su3D-result} and \ref{su3-result}. We list only the results for $SU(3)$ subgroups of type D with order $< 512$ in Table \ref{su3D-result} and six exceptional groups in Table \ref{su3-result}. The remaining groups are of type C which give only tri-maximal $|c_0|$. In the Tables, the various groups are represented by \group{$O$, $N$} following the convention used by GAP \cite{GAP4} which is a computer package that contains the groups of small order. Here, $O$ is an order of a group and $N$ indicates identification number of that group in GAP. Some common predictions for column are denoted as TM (tri-maximal), BM (bi-maximal), TL (tri-large) and 
BL (bi-large). They correspond to 
\beqa
{\rm TM}=\left( \frac{1}{3},\frac{1}{3},\frac{1}{3}\right)^T,~~{\rm BM}=\left( 0,\frac{1}{2},\frac{1}{2}\right)^T,~~{\rm TL}=\left( \frac{2}{3},\frac{1}{6},\frac{1}{6}\right)^T,~~{\rm BL} =\left( \frac{1}{2},\frac{1}{4},\frac{1}{4}\right)^T~ \eeqa
A number in the last column in tables indicates $n^{\rm th}$ column in $|U_{\rm PMNS}^{\rm exp}|$ with which the corresponding prediction is in the agreement.

\begin{table}[!ht]
\begin{small}
\begin{center}
\begin{tabular}{cccc}
 \hline
 \hline
Group & Classification  & $|c_0|^2$  & Best fit\\
 \hline
 \hline
 \group{24,12} & $\Delta(6 \times 2^2)$  & BM, BL  &  -- \\
               &   & TM  &  2 \\
               &   & TL  &  1 \\
 \hline
 \group{54,8} & $\Delta(6 \times 3^2)$  & BM  &  -- \\
 \hline
 \group{96,64} & $\Delta(6 \times 4^2)$  & BM, BL  &  -- \\
               &   & TM  &  2 \\
               &   & TL  &  1 \\
               &   & $(0.622,0.3333,0.0447)^T$, $(0.8536,0.1464,0)^T$  &  -- \\
 \hline
 \group{150,5} & $\Delta(6 \times 5^2)$  & BM, BL  &  -- \\
               &   & $(0.6594,0.2303,0.1103)^T$  &  1 \\
               &   & $(0.603,0.3682,0.0288)^T$, $(0.9045,0.0955,0)^T$  &  -- \\
               &   & $(0.6545,0.3455,0)^T$  &  -- \\
 \hline
  \group{162,14} & $D^1_{9,3}$  & BM, BL  &  -- \\
               &   & $(0.6466,0.2755,0.078)^T$  &  1 \\
               &   & $(0.75,0.25,0)^T$  &  -- \\
 \hline
 \group{216,95} & $\Delta(6 \times 6^2)$  & BM, BL  &  -- \\
               &   & TM  &  2 \\
               &   & TL  &  1 \\
               &   & $(0.933,0.067,0)^T$  &  -- \\
 \hline
  \group{294,7} & $\Delta(6 \times 7^2)$  & BM, BL  &  -- \\
               &   & $(0.6629,0.2116,0.1255)^T$  &  1 \\
               &   & $(0.5777,0.4075,0.0148)^T$, $(0.6337,0.3084,0.0579)^T$  &  -- \\
               &   & $(0.9505,0.0495,0)^T$, $(0.6113,0.3887,0)^T$  &  -- \\
               &   & $(0.8117,0.1883,0)^T$  &  -- \\
 \hline
  \group{384,568} & $\Delta(6 \times 8^2)$  & BM, BL  &  -- \\
               &   & TM  &  2 \\
               &   & TL, $(0.6553,0.2471,0.0976)^T$  &  1 \\
               &   & $(0.622,0.3333,0.0447)^T$, $(0.569,0.4196,0.0114)^T$  &  -- \\
               &   & $(0.9619,0.0381,0)^T$, $(0.6913,0.3087,0)^T$  &  -- \\
               &   & $(0.8536,0.1464,0)^T$  &  -- \\ 
 \hline
  \group{486,61} & $\Delta(6 \times 9^2)$  & BM, BL  &  -- \\
               &   & $(0.6466,0.2755,0.078)^T$  &  1 \\
               &   & $(0.9698,0.0302,0)^T$, $(0.75,0.25,0)^T$  &  -- \\
               &   & $(0.5868,0.4132,0)^T$, $(0.883,0.117,0)^T$   &  -- \\
\hline
\hline
\end{tabular}
\end{center}
\end{small}
\caption{Predictions for column $c_0$ in lepton mixing matrix from DSG of $SU(3)$ of type D and order $< 512$. The last column shows the number of a column in $|U^{\rm exp}_{\rm PMNS}|$ with which the prediction is in agreement. See the text for more details.}
\label{su3D-result}
\end{table}
\begin{table}[!ht]
\begin{small}
\begin{center}
\begin{tabular}{cccc}
 \hline
 \hline
Group & Classification  & $|c_0|^2$  & Best fit\\
 \hline
 \hline
 \group{60,5} & $A_5$, $\Sigma(60)$  & BM  &  -- \\
               &   & TM, $(0.3618,0.3618,0.2764)^T$  &  2 \\
               &   &  $(0.8727,0.0637,0.0637)^T$, $(0.4363,0.4363,0.1273)^T$ & -- \\
               &   &  $(0.7236,0.1382,0.1382)^T$ & -- \\
 \hline
 \group{108,15} & $\Sigma(36 \phi)$  & BM  &  -- \\
                &   &  $(0.5915,0.25,0.1585)^T$ & -- \\
 \hline
 \group{168,42} & $PSL(2,7)$, $\Sigma(168)$  & BM, BL  &  -- \\
               &   & TM  &  2 \\
               &   & TL  &  1 \\
               &   & $(0.664,0.2045,0.1315)^T$, $(0.7057,0.25,0.0443)^T$  &  1 \\
               &   & $(0.7985,0.1667,0.0348)^T$  &  -- \\
 \hline
  \group{216,88} & $\Sigma(72 \phi)$  & BM  &  -- \\
                &   &  $(0.5915,0.25,0.1585)^T$ & -- \\
 \hline
  \group{648,259} & $\Sigma(216 \phi) $  & BM, BL  &  -- \\
               &   & $(0.6466,0.2755,0.078)^T$  &  1 \\
               &   & $(0.5915,0.25,0.1585)^T$, $(0.75,0.25,0)^T$  &  -- \\
  \hline
    \group{1080, 260} & $\Sigma(360 \phi) $  & BM, BL  &  -- \\
               &   & TM, $(0.3618,0.3618,0.2764)^T$  &  2 \\
               &   & TL  &  1 \\
               &   & $(0.8727,0.0637,0.0637)^T$, $(0.4363,0.4363,0.1273)^T$  & -- \\
               &   & $(0.5467,0.3618,0.0915)^T$, $(0.7236,0.1382,0.1382)^T$  &  -- \\
               &   & $(0.7992,0.1382,0.0626)^T$, $(0.8026,0.1019,0.0955)^T$  &  -- \\
               &   & $(0.5915,0.25,0.1585)^T$, $(0.6545,0.3066,0.0389)^T$  &  -- \\
\hline
\hline
\end{tabular}
\end{center}
\end{small}
\caption{Predictions for column $c_0$ in lepton mixing matrix from some exceptional DSG of $SU(3)$ that are not of type C or D. The last column shows the number of a column in $|U^{\rm exp}_{\rm PMNS}|$ with which the prediction is in agreement. See the text for more details.}
\label{su3-result}
\end{table}

We make the  following observations form the results given in Table \ref{su3D-result} and \ref{su3-result}.
\begin{itemize}
\item The numerical results obtained in case of groups of type D are consistent with the analytic results derived in terms of eq.  (\ref{6nsquarepredict3}) and Fig. \ref{fig1}. In case of $\Delta(6n^2)$ groups and $n\le9$, there exist five different solutions which can reproduce the first column of $U^{\rm exp}_{\rm PMNS}$ within $3\sigma$. Using the current best fit values, $\sin^2\theta_{23}=0.440$ and $\sin^2\theta_{13}=0.02163$, obtained from the global fit \cite{Esteban:2016qun}, one can determine the predictions for solar angle and Dirac CP phase from these columns. They are given as:
\beqa \label{firstcolumns}
{\rm TL} & \Rightarrow &~~\sin^2\theta_{12}= 0.319,~~\cos\delta=-0.263   ~~\nonumber \\
(0.6594,0.2303,0.1103)^T & \Rightarrow &~~\sin^2\theta_{12}= 0.326,~~\cos\delta=0.607  ~~\nonumber \\
(0.6466,0.2755,0.078)^T & \Rightarrow &~~\sin^2\theta_{12}= 0.339,~~\cos\delta \to {\rm inconsistent}   ~~\nonumber \\
(0.6629,0.2116,0.1255)^T & \Rightarrow &~~\sin^2\theta_{12}= 0.322,~~\cos\delta=0.364  ~~\nonumber \\
(0.6553,0.2471,0.0976)^T& \Rightarrow &~~\sin^2\theta_{12}= 0.330,~~\cos\delta=0.817 \eeqa
One column in the above gives $\cos\delta>1$ which is inconsistent, however these results change when deviations in $\theta_{13}$ and $\theta_{23}$ from their best fit values are considered. The group $D^1_{9,3}$ gives same results as $\Delta(6 \times 9^2)$. No prediction consistent with the second column of $U^{\rm exp}_{\rm PMNS}$ other than TM exists as shown analytically in the previous section.
\item The DSG of $SU(3)$ which are not of type C or D lead to a new solution for the second column other than TM. The predicted column $(0.2764, 0.3618, 0.3618)^T$ implies
\be
\sin^2\theta_{12} = 0.283~~~{\rm and}~~~\cos\delta=0.638
\ee
given the best fit values of $\theta_{23}$ and $\theta_{13}$. This prediction is offered by the group $A_5$ and the group $\Sigma(360 \phi)$ which contains $A_5$ as a subgroup.
\item The group $PSL(2,7)$ or $\Sigma(168)$ provides two new solutions consistent with the first column of $U^{\rm exp}_{\rm PMNS}$. They imply
\beqa \label{firstcolumns-psl27}
(0.664,0.2045,0.1315)^T & \Rightarrow &~~\sin^2\theta_{12}= 0.321,~~\cos\delta =0.270 ~~\nonumber \\
(0.7057,0.25,0.0443)^T& \Rightarrow &~~\sin^2\theta_{12}= 0.279,~~\cos\delta \to {\rm inconsistent}  \eeqa
for the best fit values of $\theta_{23}$ and $\theta_{13}$. Predictions for this group were also analyzed in \cite{Hernandez:2012sk} where a solution for the second column leading to correct
$\sin^2\theta_{12}$ was found. Prediction of the entire column obtained in this case
does not however reproduce the entire second column consistent with eq. (\ref{expUPMNS}). The group $\Sigma(216 \phi)$ gives a solution consistent with the first column which can also be obtained from  $\Delta(486)$ and hence also a type D group $D^1_{9,3}$. 
\item No viable prediction for the third column can be obtained from the DSF of $SU(3)$ with order $< 512$.
\end{itemize}

The results obtained for the groups in Table \ref{su3-result} together with analytic results obtained for the groups of type C and D exhaust all the results that can be obtained from DSG of $SU(3)$ possessing three dimensional IR. As can be seen from Fig. \ref{fig1}, the smallest group reproducing consistent values for the third column is a group $\Delta(6n^2)$ with $n=11$. Interestingly for $n=10$, one finds a fixed column $|c_0|^2=(0.603,0.3682,0.0288)^T$ implying $\sin^2 \theta_{13} = 0.0288$ which is larger than the current value allowed at the $3\sigma$ by the latest global fit NuFIT 3.0 \cite{Esteban:2016qun}. The group $\Delta(6 \times 10^2)$ was identified in \cite{Holthausen:2012wt,Yao:2015dwa} as the smallest group predicting all the three neutrino mixing angles within their $3\sigma$ ranges implied by the global fits of that time. If the results from latest global fit NuFIT 3.0 \cite{Esteban:2016qun} are considered then this conclusion gets modified and $\Delta(6 \times 11^2)$ becomes the smallest 
group which reproduces consistent $\sin^2\theta_{13}$. The predicted value, $\sin^2\theta_{13}=0.0239$, however is just below the maximum value allowed at $3\sigma$ by the current fit. The group $\Delta(6 \times 11^2)$ also gives two different viable solutions for the first column of $U_{\rm PMNS}$. However, it cannot reproduce both the first and third columns simultaneously as the $Z_2$ symmetries generating these two different solutions do not commute with each others. Since the order of $\Delta(6 \times 11^2)$ is not divisible by 4, it cannot contain two mutually commuting $Z_2$ groups and hence cannot lead to the correct predictions for any two columns simultaneously. An another arguments is that the group $\Delta(6 \times 11^2)$ corresponding to odd value of $n$ does not contain the tri-maximal solution which is the only viable solution for the second column of $U_{\rm PMNS}$ offered by $\Delta(6n^2)$ groups. Therefore, $\Delta(6 n^2)$ with odd $n$ cannot lead to the viable predictions for all the tree 
mixing angles simultaneously.
As  seen from Fig. \ref{fig1}, the smallest even $n$ reproducing a consistent solution for the third column is $n=18$ corresponding to the group of order 1944 \footnote{In \cite{Holthausen:2012wt,Yao:2015dwa}, the group $\Delta(6 \times 16^2)$ was identified as the smallest viable group reproducing all the mixing angles within the $3\sigma$ ranges of their global fit values. Again, it predicts $\sin^2\theta = 0.0254$ which is outside the $3\sigma$ range implied by the latest global fit considered by us.}. We get the following elements in $\Delta(6 \times 18^2)$ which can be identified as the $Z_2\times Z_2$ symmetry of neutrinos and $Z_3$ symmetry of the charged leptons.
\be 
S_{1\nu} \equiv GB= - \left(
\begin{array}{ccc}
 1& 0 & 0 \\
 0 & 0 & \epsilon^* \\
 0 & \epsilon & 0 \\
\end{array}
\right),~~S_{2 \nu} \equiv G^9 = W(1,-1,-1),~~T_l=E, \ee
where the generators $G$, $B$ and $E$ are given in eq. (\ref{generators}) with $\epsilon = e^{2 \pi i/18}$. The $S_{1\nu, 2\nu}$ are diagonalized by appropriate $V_S$ given in eq.  (\ref{ug}) while $T_l$ is diagonalized by $U_\omega$ given in eq.  (\ref{uw}). Following eq.  (\ref{PMNS}) and rearranging columns, they lead to
\be
|U_{\rm PMNS}| = \left(
\begin{array}{ccc}
 0.804 & 0.577 & 0.142 \\
 0.525 & 0.577 & 0.625 \\
 0.279 & 0.577 & 0.767 \\
\end{array}
\right)\ee
which is consistent with $U^{\rm exp}_{\rm PMNS}$ given in eq.  (\ref{expUPMNS}). It predicts
\be \label{1944predictions}
\sin^2\theta_{12}=0.340,~~\sin^2\theta_{23}=0.399~{\rm or}~0.601,~~\sin^2\theta_{13}=0.0201,~~\cos\delta=1~. \ee
The same predictions can also be obtained from the group $D^1_{9n',3n'}$ with $9n'=n=18$ as shown analytically in the previous section. The group $D^1_{18,6}$ has the rank 648 and therefore is the smallest group predicting the viable values for all the three mixing angles.

Let us now consider DSG of $U(3)$ which are not subgroups of $SU(3)$. There exist 75 such groups with three dimensional IR and order $<512$. As can be seen from Table 5 in \cite{Ludl:2010bj}, most of these group are generated by a texture of type $R(\eta_1,\eta_2,\eta_3)$ given in eq.  (\ref{textures}). All such groups have elements of type $W$, $R$ and $V$. Among these, only $W$ can generate $Z_2$ subgroups leading to tri-maximal $|c_0|$ and predictions eq.  (\ref{tri-maximal}) and this is the only solution possible for all such groups. We therefore give results for the  remaining groups in Table \ref{u3-result} which are generated by the textures of type $S$, $T$, $U$ or $X_i$ with $i=1,2,...,10$.
\begin{table}[!ht]
\begin{small}
\begin{center}
\begin{tabular}{cccc}
 \hline
 \hline
Group & ~~~~~~Classification~~~~~~  & ~~~~$|c_0|^2$~~~~   & Best fit\\
 \hline
 \hline
\group{48,30} & $S_4(2)$   & BM & -- \\
					     &	 & TM & 2\\
\hline					     	     
\group{96,65} & $S_4(3)$   & BM & -- \\
					     &	 & TM & 2\\
\hline
\group{162,10} &    & BM, BL & -- \\
				    &	 & $(0.6466,0.2755,0.078)^T$ & 1\\
					&	 & $(0.75,0.25,0)^T$ & 0\\
\hline
\group{162,12} &    & BM, BL & -- \\
				    &	 & $(0.6466,0.2755,0.078)^T$ & 1\\
					&	 & $(0.75,0.25,0)^T$ & 0\\
\hline	
\group{162,44} & $\Delta(6\times 3^2,2,1)$  & BM & -- \\
\hline					     	     
\group{192,186} & $S_4(4)$   & BM & -- \\
					     &	 & TM & 2\\			     
\hline	
\group{324,111} &    & BM, $(0.5915,0.25,0.1585)^T$ & -- \\	     
\hline					     	     
\group{384,571} & $\Delta(6\times 4^2,3)$   & BM & -- \\
					     &	 & TM & 2\\			     
\hline					     	     
\group{384,581} & $S_4(5)$   & BM & -- \\
					     &	 & TM & 2\\			     
\hline	
\group{432,239} &    & BM, $(0.5915,0.25,0.1585)^T$ & -- \\	     
\hline	    
\group{432,260} & $\Delta(6\times 4^2,2)$   & BM & -- \\
					     &	 & TM & 2\\			     
\hline	
\group{432,273} &    & BM, $(0.5915,0.25,0.1585)^T$ & -- \\	     
\hline
\group{486,26} &    & BM, BL & -- \\
				    &	 & $(0.6466,0.2755,0.078)^T$ & 1\\
					&	 & $(0.75,0.25,0)^T$ & 0\\
\hline	
\group{486,28} &    & BM, BL & -- \\
				    &	 & $(0.6466,0.2755,0.078)^T$ & 1\\
					&	 & $(0.75,0.25,0)^T$ & 0\\
\hline	
\group{486,125} &    & BM, BL & -- \\
				    &	 & $(0.6466,0.2755,0.078)^T$ & 1\\
					&	 & $(0.75,0.25,0)^T$ & 0\\
\hline		
\group{486,164} & $\Delta(6\times 3^2,3,1)$  & BM & -- \\
\hline
\hline
\end{tabular}
\end{center}
\end{small}
\caption{Predictions for column $c_0$ in lepton mixing matrix from DSG of $U(3)$ with order $<512$. The last column shows the number of a column in $|U^{\rm exp}_{\rm PMNS}|$ with which the prediction is in agreement. See the text for more details.}
\label{u3-result}
\end{table}
None of these groups generate  any new mixing patterns compared to ones already found with the DSG of $SU(3)$. This  result is  anticipated from our analytic discussion in the previous section for all the groups which do not involve exceptional generators $X_i$. All the DSG of $U(3)$ of order $< 512$ lead to a column of type TM or $(0.6466,0.2755,0.078)^T$ which can be identified with the second and first column of $U_{\rm PMNS}$ respectively.

Before ending this section, we make a comparison of the results found by us with the previous similar studies. In \cite{Lam:2012ga}, the DSG of $SU(3)$ with order up to 511 were numerically searched for their column prediction. Our analysis includes \emph{all} the DSG of $SU(3)$ and also DSG of $U(3)$ which are not subgroups of $SU(3)$. An analytical treatment of $SU(3)$ subgroups of type C and D allows us to extend our results to larger groups. As a result of this, we find groups which can predict viable third column for $U_{\rm PMNS}$ unlike \cite{Lam:2012ga}\footnote{The group $\Delta(6\times 5^2)$ generates a solution $|c_0|^2 = (0.0289, 0.368, 0.604)^T$ which was identified in \cite{Lam:2012ga} as a consistent prediction for the third column of $U_{\rm PMNS}$. However, it predicts $\sin^2\theta_{13}$ greater than the highest value allowed by the current global fit at $3\sigma$.}. Our results for the $SU(3)$ subgroups of order $< 512$ matches with their results. Ref. \cite{Yao:2015dwa} mainly considered 
Dirac neutrinos but also included a discussion of Majorana neutrinos. Their numerical scan of groups of order $<2000$ looked at the Klein groups and identified 10 groups giving consistent patterns. All these predict vanishing Dirac CP phase. Use of the only $Z_2$ groups allows non-trivial prediction for this phases as  is apparent, for example, from eq. (\ref{firstcolumns}). Our results provide  generalization of the existing results in which prediction for one or more columns of PMNS matrix were obtained using residual $Z_2$ or $Z_2 \times Z_2$ symmetries. Most importantly, we analytically and numerically show that the groups from $\Delta(6n^2)$ series are sufficient and sometime are the smallest groups leading to specific prediction for a column of PMNS matrix which follows from the groups of type C, D and some known group series representing DSG of $U(3)$.

\section{Additional symmetries}
\label{additional}
The $U_{\rm PMNS}$ matrix of eq. (\ref{unuul}) derived from the previous considerations contain the unknown phases and angle $\theta$ which can affect the observables. Some or all of these can be fixed  by imposing additional symmetries. These have been discussed analytically \cite{Yao:2015dwa} for the type C and D groups and also
have been numerically explored.  We show that essentially the same analysis gets carried over also to the DSG of $U(3)$ being studied here. We discuss here only  the case giving non-zero values  for all the entries of $\Upmns$ corresponding to $S_\nu$ in (\ref{s}) and $T_l$ having the texture of $R,V$. Other cases are not much of phenomenological interest but can be discussed analogously.
The $U_{\rm PMNS}$  in this case can explicitly be written as
\be \label{upmnsexp}
U_{\rm PMNS}=\left(\ba{ccc}
\frac{\eta_1\Delta^*}{\sqrt{6}}(1-\eta_\nu\eta_2\Delta^*)&\frac{1}{\sqrt{3}} &\frac{\eta_1 \Delta^*}{\sqrt{6}}(1+\eta_\nu\eta_2\Delta^*)\\
\frac{\eta_1\omega \Delta^*}{\sqrt{6}}(1-\omega\eta_\nu\eta_2\Delta^*)& \frac{1}{\sqrt{3}} & \frac{\eta_1\omega \Delta^*}{\sqrt{6}}(1+\omega\eta_\nu\eta_2\Delta^*)\\
\frac{\eta_1\omega^2 \Delta^*}{\sqrt{6}}(1-\omega^2\eta_\nu\eta_2\Delta^*) &\frac{1}{\sqrt{3}}& \frac{\eta_1\omega^2 \Delta^*}{\sqrt{6}}(1+\omega^2\eta_\nu\eta_2\Delta^*)\\
\ea \right) P \left(\ba{ccc}
1&0&\\
0&\cos\theta&-\sin\theta e^{i\beta}\\
0&\sin\theta e^{-i \beta}&\cos\theta\\ \ea\right).\ee

The first column of the above matrix is $\theta$ independent and already given in eq. (\ref{predict3}).
$P$ is a diagonal phase matrix with $P_{jj}=e^{i\chi_j}$. The $\theta$ dependence can be eliminated by imposing another $Z_2$ symmetry commuting with the first $Z_2$ defined by 
$S_{1\nu}\equiv S_\nu$ as given in eq. (\ref{s}). The most general $3\times 3$ matrix having the texture in eq. (\ref{textures}) and satisfying $[S_{1\nu},S_{2\nu}] =0$, $S_{2\nu}^2={\bf 1}$ and ${\rm det.}(S_{2\nu})=1$ 
is either $W(1,-1,-1)={\rm diag.}(1,-1,-1)$ or
\be \label{s2}
S_{2\nu}=\left( \ba{ccc}
-1&0&0\\
0&0&\eta_\nu^*\\
0&\eta_\nu&0\\ \ea \right)~.\ee 
The matrices ${\cal Z}\equiv \{ W(1,-1-1),S_{1\nu},S_{2\nu} \}$ together with the identity matrix define  $Z_2\times Z_2$ groups. 
These and their cyclic permutations  define all possible Klein groups  having non-diagonal structures in DSG of $SU(3)$ and $U(3)$ being considered here. 
Since $\eta_\nu$ represents $n^{\rm th}$ root of identity for some $n$,  both $S_{1\nu}$ and $S_{2\nu}$ can be present in the same group only for even $n$ and the residual Kelin symmetry can be imposed only in such groups.
The set ${\cal Z}$  is simultaneously diagonalized by
\be \label{vnu}
V_\nu=\frac{1}{\sqrt{2}}\left(\ba{ccc}
0&\sqrt{2}&0\\
1&0&1\\
-\eta_\nu&0&\eta_\nu\\ \ea \right)\ee
$$V_\nu^\dagger S_{1\nu}V_\nu={\rm diag.}(1,-1,-1)~,~~~ V_\nu^\dagger S_{2\nu}V_\nu={\rm diag.}(-1,-1,1)~.$$
This $V_\nu$ with $ V_l=V_{R,S}$ defined in eq. (\ref{ug}) give the $U_{\rm PMNS}$ as in eq. (\ref{upmnsexp}) but now $\theta$ is fixed to be zero allowing group theoretical prediction for $|U_{\rm PMNS}|$. Eq. (\ref{upmnsexp}) with $\theta=0$ is the most general $U_{\rm PMNS}$ possible in all the groups under consideration when $T_l$ has the structure $R$ or $V$.  An alternative and phenomenologically less interesting possibility for the $Z_2\times Z_2$ groups is a set of diagonal matrices. This gives democratic structure for $|\Upmns|$.
\begin{itemize}
\item Eq. (\ref{upmnsexp}) with $\theta=0$ has been derived earlier for the groups $\Delta(6 n^2)$ \cite{King:2013vna} as well as all  type D groups \cite{Yao:2015dwa}. The derivation here shows that it has more general validity and holds for all the groups having elements with textures as in eq. (\ref{textures}) and containing the Klein group. Only change is the occurrence of the factor $\Delta$ which is defined as $\Delta^3= {\rm det.}(T_l)$. In particular, the general result that one of the columns of $|U_{\rm PMNS}|$ necessarily has  tri-maximal form  if entries in other two columns are required to be non-zero remains true here as well.
\item For the $U(3)$ group series considered here the root  $\eta_2$ differs from a power of the $n^{th}$ root of unity only by an overall factor equal to $\Delta$ with the result that eq. (\ref{upmnsexp}) coincide with the corresponding result for $\Delta(6 n^2)$. 
As already discussed only group series having $\Delta(6 n^2)$ as subgroup is $\Delta(6 n^2,j,1)$. In that case the Klein groups in the latter coincide with the ones used in the above derivation and the use of $\Delta(6 n^2,j,1)$ instead of $\Delta(6 n^2)$ does not give any new result. The remaining group series of $U(3)$ can contain only diagonal Klein groups and give only democratic mixing.
\item Eq. (\ref{upmnsexp}) leads to vanishing Jarlskog invariant \cite{Olive:2016xmw} for $\theta=0$ as can been explicitly verified. Thus the Dirac CP violation is absent in very large class of models containing the Klein group. This result changes if one only requires $Z_2$ symmetry which does not fix $\theta$. We give an explicit example below. 
\end{itemize}

Eq. (\ref{upmnsexp}) reduces to the tri-bimaximal form in  special cases of $\eta_2\eta_\nu\Delta^*=(-1,-\omega,-\omega^2)$. This pattern thus gets realized in large number of groups (e.g. all the $\Delta(6 n^2)$ groups with even $n$) as the derivation here shows. The required departures from the tri-bimaximal mixing can be generated by using only a single $Z_2$ symmetry generated by $S_{1\nu}$. In that case the first column has the desired absolute values $(\sqrt{2/3}, 1/\sqrt{3}, 1/\sqrt{3})^T$ but the other two columns are $\theta$ dependent.  Even in that case,  the invariants given in eq. (\ref{Aalpha}) are $\theta$ independent if $\beta=0$ is assumed. Using eq. (\ref{upmnsexp}) one now gets
\beqa \label{invariants1}
A_e&=&0=c_{13}s_{13}s_{12}\sin(\delta+\beta_2-\beta_3)~,\nonumber \\
A_\mu&=&-\frac{1}{\sqrt{6}}~ {\rm Re}\left(\eta_1\Delta^*e^{i(\chi_2-\chi_3)}\right)~\nonumber \\
   &=&c_{13}c_{12}c_{23}s_{23}\sin(\beta_2-\beta_3)-
c_{13}s_{13}s_{12}s_{23}^2\sin(\delta+\beta_2-\beta_3)~.\eeqa
Here the second entry in each of these equations  is obtained from eq. (\ref{upmnsexp}) while the last entry from the standard parameterization of the PMNS matrix \cite{Olive:2016xmw} with $\beta_2$, $\beta_3$ denoting the Majorana phases.  From now on, we specialize our discussions to groups $\Delta(6 n^2)$ which as argued earlier are sufficient to 
describe mixing patterns of all other groups. In that case, eq. (\ref{invariants1}) imply
\beqa \label{invariants2}
\beta_2-\beta_3+\delta&=&k \pi ~,\nonumber\\
c_{23}s_{23}\sin\delta&=&\pm\frac{1}{2}~{\rm Re}\left(\eta_1e^{i(\chi_2-\chi_3)}\right)~.\eeqa
There still remains dependence on the arbitrary phases $\chi_{2,3}$. This can be fixed by invoking CP
as a residual symmetry as has been discussed in variety of ways \cite{Grimus:2003yn,Kitabayashi:2005fc,Farzan:2006vj,Joshipura:2009tg,Ge:2011ih,Gupta:2011ct,Grimus:2012hu,Mohapatra:2012tb,Feruglio:2012cw,Holthausen:2012dk,Chen:2014tpa,King:2014rwa,Hagedorn:2014wha,Ding:2014ora,Ding:2015rwa,Yao:2016zev}\footnote{There is an extensive literature on the subject and we refer to \cite{Yao:2016zev} for a fairly exhaustive list of references.}. We give a simple example here based on $\Delta(6 n^2)$ with CP as residual symmetry of neutrinos. A more detailed analysis of such case is carried out in \cite{Ding:2014ora}.
Assume that $M_\nu$ satisfies eq. (\ref{invariance}) with $S_\nu=S_{1\nu}$ and transforms as follows  under a CP symmetry $X_\nu$
\be \label{cp} X_\nu^TM_\nu X_\nu=M_\nu^*~,\ee
where $X_\nu$ denotes a $3\times 3$ matrix corresponding to the $\mu$-$\tau$ interchange symmetry. The $X_\nu$ and $S_{1\nu}$ satisfy $X_\nu S_{1\nu}^*-S_{1\nu} X_\nu=0$ as required  to
define the direct product of $Z_2$ and CP as a consistent residual symmetry \cite{Feruglio:2012cw}. The most general $M_\nu$ satisfying eqs. (\ref{invariance},\ref{cp}) has the form
\be \label{mnu}
M_\nu=\left(
\ba{ccc}
x&a\eta_\nu^{\frac{1}{2}}&a\eta_\nu^{-\frac{1}{2}}\\
a\eta_\nu^{\frac{1}{2}}&b\eta_\nu&c\\
a\eta_\nu^{-\frac{1}{2}}&c&b\eta_\nu^*\\ \ea \right)\ee
where $x,a,b,c$ are real. This $M_\nu$ is diagonalized by 
$$U_\nu=V_\nu ~{\rm diag.}(\eta_\nu^{-\frac{1}{2}},1,\eta_\nu^{-\frac{1}{2}})R_{23}(\theta)$$
with real rotation $R_{23}(\theta)$. This implies $\beta=0$ and $e^{i(\chi_2-\chi_3)}=\eta_\nu^{1/2}$. Eq. (\ref{invariants2}) now gives for the $\Delta(6 n^2)$ groups 
\be \label{delta1}
c_{23}s_{23}\sin\delta=\pm \frac{1}{2} \cos\left(\frac{2\pi\left(q+\frac{q_\nu}{2}\right)}{n}\right)~,\ee
where we have defined $\eta_1=e^{2\pi i q/n},\eta_\nu=e^{2\pi i q_\nu/n}$. This equation is $\theta$ independent but $c_{23} s_{23}$ depends on $\theta$ implicitly. One however has the following $\theta$ independent correlation
$$s_{23}^2=\left[\frac{1}{2}\pm \frac{\sqrt{2}s_{13}\sqrt{1-3 s_{13}^2}}{c_{13}^2} ~{\rm Im}(\eta_1)\right]^{\frac{1}{2}}.$$
$\theta_{23}$ is thus predicted to be nearly maximal and eq. (\ref{delta1}) is now approximately given by
\be \label{delta2}
\sin\delta\approx \pm \cos \left(\frac{2\pi\left(q+\frac{q_\nu}{2}\right)}{n}\right)~.\ee
It is seen that all the $\Delta(6n^2)$ groups giving identical prediction $(\sqrt{2/3}, 1/\sqrt{3}, 1/\sqrt{3})^T$ for the first column differ in their prediction for the CP phases. The minimal case corresponding to $n=2$, namely the $S_4$ group gives $\delta=0, \pm \frac{\pi}{2},\pm \pi$,
while $n=4$ corresponding to $\Delta(96)$ predict $\delta\approx (0,\pm 45^\degree,\pm 90^\degree)$. 
In general, for $n=2k$, $\delta\approx \pm \frac{\pi l}{2k}$ and $l=0,1,...,k$.

\section{Summary}
\label{summary}
Discrete subgroups $G_f$ of $SU(3)$ as flavour symmetries of the leptonic world have been  extensively discussed in the last decade. The possible mixing patterns implied by these groups have been studied both analytically and numerically. We have given here an analytic discussion of the mixing angle predictions within the DSG of $U(3)$  assuming that neutrino mass matrix is invariant under a single $Z_2$ group $\in G_f$. The resulting mixing formula eqs. (\ref{predict1}-\ref{predict5}) encompasses predictions of large number of groups. 

The consequences of assuming a single $Z_2$ symmetry as a residual symmetry of $M_\nu$ have been already discussed in variety of cases. Ref. \cite{Hernandez:2012sk,Hernandez:2012ra} discussed
predictions of the finite and some of the infinite subgroups of the von-Dyck groups. These
predictions have also been studied  for the DSG of $SU(3)$ of  C and D type  which emerge here as special cases of our general formula. Variant of these have also been studied numerically in \cite{Lavoura:2014kwa,Jurciukonis:2016wrh}.
 
Most of the above studies concentrated on the DSG of $SU(3)$. We have also discussed here symmetries which are contained in $U(3)$ but not in $SU(3)$. Not all DSG of $U(3)$ are known but a large number of them have elements with textures given in eq. (\ref{textures}). The mixing angle predictions do not directly involve basic structure of the groups and use only these textures. This makes them useful to study many groups. One can group theoretically predict absolute magnitude of  any one column of the mixing matrix $\Upmns$ under the assumption of single $Z_2$ symmetry. We have analytically studied this predictions for all the known series of groups which are contained in $U(3)$ using the  structures derived here. One of the  important conclusion of this work is the realization that the prediction of a single column of $|\Upmns|$ in all these group series can be obtained using $\Delta(6n^2)$ as a flavour symmetry for some appropriate $n$. This conclusion applies to group series of types C, D and various 
series contained in $U(3)$ \cite{Ludl:2010bj}. Specifically, (a) the group series $D^1_{9 n',3 n'}$ lead to the same mixing pattern as the groups $\Delta(486 n'^2)$, (b) the series $\Sigma(3 n^3)$, $\Delta(3 n^2,j)$ for even $n$ lead to tri-maximal pattern as implied by $\Delta(3 n^2)$ contained in $\Delta(6n^2)$, (c) The $\Delta(6 n^2,m)$ groups lead to the tri-maximal and the bi-maximal patterns, and (d) mixing in case of the $\Delta'( 6n^2,j,k)$ groups coincide with that in $\Delta(6 n^2)$ for $k=1$ and with tri-maximal and bi-maximal for other $k$. All in all, the group series $\Delta(6 n^2)$  represent a sufficient set of discrete groups for the study of predictions of a single $Z_2$ residual symmetry. 

The above analytic results are supplemented with numerical discussion using the latest global fit of the neutrino mixing angles. It is found that some of the earlier numerical results get modified in view of the latest result and earlier allowed cases such as $\Delta(150)$, $\Delta(600)$ do not give results consistent with the present 3$\sigma$ ranges of the mixing angles. The groups $\Delta(6n^2)$ (and consequently the large class of DSG of $U(3)$ as discussed above) lead to only tri-maximal mixing as the viable solution for the second column of leptonic mixing matrix. These groups can predict either first or third column of $|\Upmns|$ for some values of $n$. The smallest group which can accommodate the third column of $U_{\rm PMNS}$ matrix within 3$\sigma$ is found to be $\Delta(6 \times 11^2)$. 
The group $A_5$ which does not belong to type C or D provides a viable solution for the second column alternative to the tri-maximal. Selections of two commuting $Z_2$ symmetries as the symmetry of neutrino mass matrix within a given group $G_f$ is known for fixing $|U_{\rm PMNS}|$ entirely. The smallest such group  in the $\Delta(6 n^2)$ series which can simultaneously predict all the three mixing angles within their 3$\sigma$ ranges is $\Delta(6 \times 18^2)$ of order 1944. The same mixing pattern also follows from the order 648 group $D_{18,6}^1$ which represents the smallest possible group containing residual symmetries leading to all the mixing angles within their 3$\sigma$ ranges.

\acknowledgments
We thank referee for pointing out a general connection between the predictions of DSG of $U(3)$ and $SU(3)$ in case of the residual symmetries considered in this paper. Work of ASJ was supported by BRNS (DAE) and by Department of Science and Technology, Government of India through the Raja Ramanna  and the J. C. Bose grant respectively. KMP thanks the Department of Science and Technology, Government of India for research grant support under INSPIRE Faculty Award (DST/INSPIRE/04/2015/000508). 

\bibliography{references}
\bibliographystyle{JHEP.bst}
\end{document}